\documentclass[aps,pra,twocolumn,amsmath,amssymb,superscriptaddress,letterpaper,longbibliography,floatfix]{revtex4-1}

\usepackage{graphicx}       
\usepackage{times}          

\usepackage[pdftex]{hyperref} 

\newcommand{\ket}[1]{|#1\rangle}
\newcommand{\bra}[1]{\langle#1|}
\newcommand{\braket}[1]{\langle #1 \rangle}
\newcommand{\abs}[1]{\lvert#1\rvert}

\begin{document}

\title{Qubit metrology of ultralow phase noise using randomized benchmarking}

\author{P. J. J. O'Malley}
\thanks{These authors contributed equally to this work}
\affiliation{Department of Physics, University of California, Santa Barbara, CA 93106, USA}
\author{J. Kelly}
\thanks{These authors contributed equally to this work}
\author{R. Barends}
\thanks{These authors contributed equally to this work}
\affiliation{Department of Physics, University of California, Santa Barbara, CA 93106, USA}
\affiliation{Present address: Google Inc., Santa Barbara, CA 93117, USA}
\author{B. Campbell}
\affiliation{Department of Physics, University of California, Santa Barbara, CA 93106, USA}
\author{Y. Chen}
\affiliation{Department of Physics, University of California, Santa Barbara, CA 93106, USA}
\affiliation{Present address: Google Inc., Santa Barbara, CA 93117, USA}
\author{Z. Chen}
\author{B. Chiaro}
\author{A. Dunsworth}
\affiliation{Department of Physics, University of California, Santa Barbara, CA 93106, USA}
\author{A. G. Fowler}
\affiliation{Department of Physics, University of California, Santa Barbara, CA 93106, USA}
\affiliation{Centre for Quantum Computation
and Communication Technology, School of Physics, The University of
Melbourne, Victoria 3010, Australia}
\affiliation{Present address: Google Inc., Santa Barbara, CA 93117, USA}
\author{I.-C. Hoi}
\affiliation{Department of Physics, University of California, Santa Barbara, CA 93106, USA}
\author{E. Jeffrey}
\affiliation{Department of Physics, University of California, Santa Barbara, CA 93106, USA}
\affiliation{Present address: Google Inc., Santa Barbara, CA 93117, USA}
\author{A. Megrant}
\affiliation{Department of Physics, University of California, Santa Barbara, CA 93106, USA}
\affiliation{Department of Materials, University of California, Santa Barbara, CA 93106, USA}
\author{J. Mutus}
\affiliation{Department of Physics, University of California, Santa Barbara, CA 93106, USA}
\affiliation{Present address: Google Inc., Santa Barbara, CA 93117, USA}
\author{C. Neill}
\author{C. Quintana}
\affiliation{Department of Physics, University of California, Santa Barbara, CA 93106, USA}
\author{P. Roushan}
\author{D. Sank}
\affiliation{Department of Physics, University of California, Santa Barbara, CA 93106, USA}
\affiliation{Present address: Google Inc., Santa Barbara, CA 93117, USA}
\author{A. Vainsencher}
\author{J. Wenner}
\author{T. C. White}
\affiliation{Department of Physics, University of California, Santa Barbara, CA 93106, USA}
\author{A. N. Korotkov}
\affiliation{Department of Electrical and Computer Engineering, University of California, Riverside, CA 92521, USA}
\author{A. N. Cleland}
\affiliation{Department of Physics, University of California, Santa Barbara, CA 93106, USA}
\author{John M. Martinis}
\affiliation{Department of Physics, University of California, Santa Barbara, CA 93106, USA}
\affiliation{Present address: Google Inc., Santa Barbara, CA 93117, USA}

\begin{abstract}
A precise measurement of dephasing over a range of timescales is critical for improving quantum gates beyond the error correction threshold.
We present a metrological tool, based on randomized benchmarking, capable of greatly increasing the precision of Ramsey and spin echo sequences by the repeated but incoherent addition of phase noise.
We find our SQUID-based qubit is not limited by $1/f$ flux noise at short timescales, but instead observe a telegraph noise mechanism that is not amenable to study with standard measurement techniques. 
\end{abstract}

\maketitle

\section{Introduction}

One of the main challenges in quantum information is maintaining precise control over the phase of a superposition state.
Long-term phase stability is threatened by frequency drifts due to non-Markovian noise, which arises naturally in solid-state quantum systems \cite{Wellstood1987,GardinerZoller}.
Fortunately, correlated noise can be suppressed using Hahn spin echo \cite{Hahn1950}.
In practice, Ramsey and spin echo measurements of dephasing \cite{Cottet2002,Biercuk2009,Bylander2011} characterize the dominant noise source for large error rates (0.1 to 0.5) and long times, but are fundamentally inappropriate for understanding noise dominant on the timescales and error rates needed for fault-tolerant gate operations ($<10^{-2}$).

In this article, we introduce a metrological tool based on randomized benchmarking (RB) \cite{Knill2008,Ryan2009,Magesan2011,Brown2011,Gambetta2012,Corcoles2013} to quantify noise on timescales relevant for quantum gates.
Whereas other measurement techniques based on Ramsey \cite{Cottet2002,Biercuk2009,Bylander2011} and Rabi \cite{Yan2013} measurements measure noise over long timescales and filter low frequency noise to infer gate performance at short timescales, we measure gate fidelity directly, providing immediate feedback on the impact of noise on gate performance.
We apply it on a superconducting quantum interference device- (SQUID-)based qubit, and show that this method determines that $1/f$ flux noise \cite{Wellstood1987,Sank2012,Yan2012} is not currently a limiting factor in our device.
This tool also provides a powerful probe of anomalous telegraph noise sources seen in superconducting devices.
We also show that undesired coherent interactions can be understood as an effective correlated noise.
Finally, we demonstrate how this method allows for error budgeting and direct selection of ideal gate parameters in the presence of non-Markovian noise.

\begin{figure}[b!]
  \centering
  \includegraphics[width=0.48\textwidth]{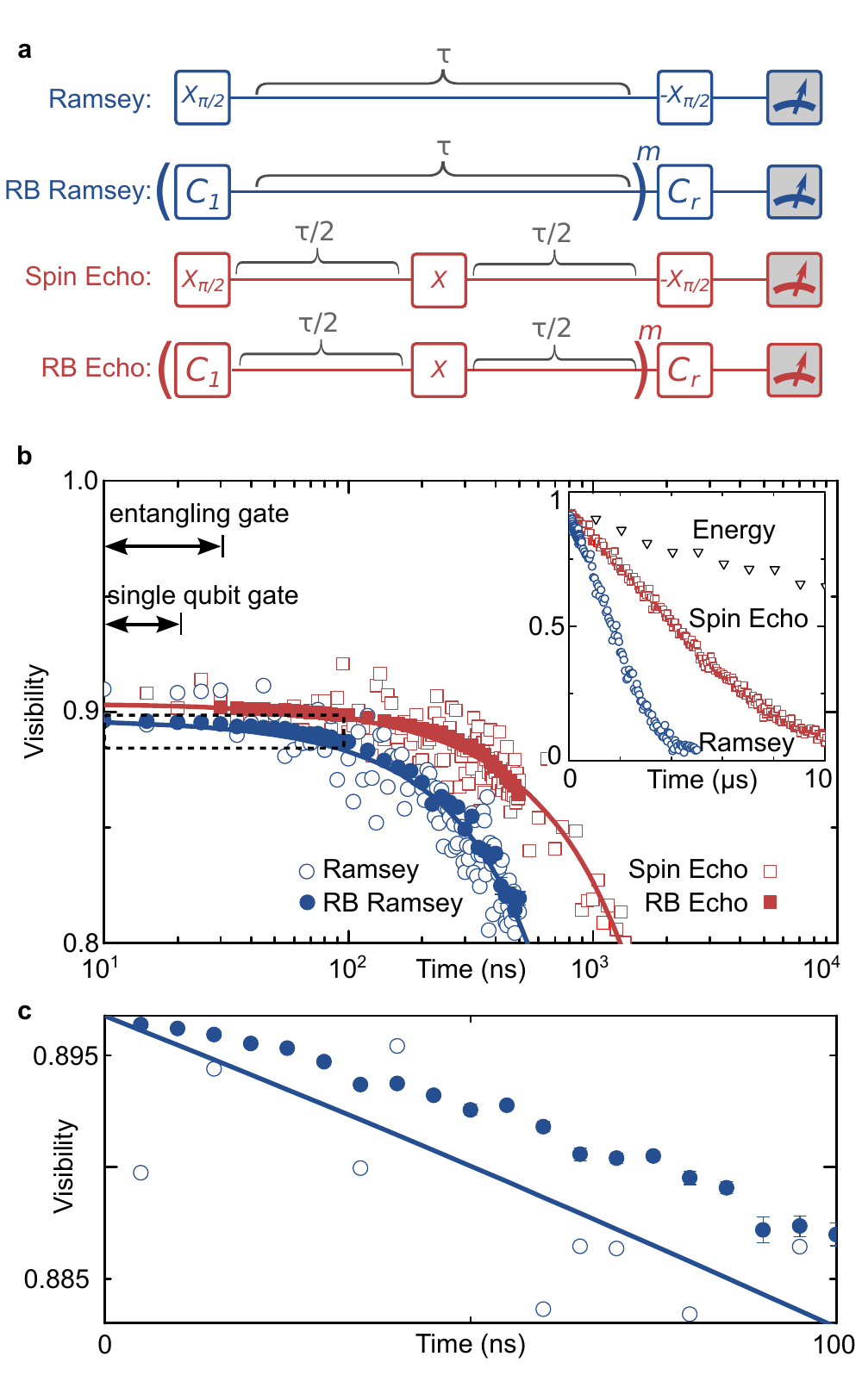}
  \caption{(color online) 
  (a) Gate diagram for Ramsey and Hahn spin echo sequences, and their RB equivalents.
  For RB Ramsey, instead of inserting an idle between $X_{\pi/2}$ pulses, we interleave the idle between $m$ randomly selected single-qubit Clifford gates ($C_1$), after which the qubit is rotated back ($C_r$) to the pole and measured.
  For spin echo and RB echo, an $X$ gate is inserted at the center of the idle.
  The range of $m$ is 21 for the longest $\tau$ to 300 for the shortest.
  (b) (inset) $T_1$ (energy decay), Ramsey, and spin echo envelopes. (main) Ramsey (open circle) and spin echo (open square) envelopes at short times.
  RB decay envelopes are inferred from $\langle \phi^2(\tau) \rangle$ measured by RB Ramsey (solid circle) and RB echo (solid square); see text for details.
  Single qubit and entangling gate durations are shown for reference.
  Note the significantly lower noise of the RB sequences, which take approximately the same measurement time as the Ramsey and echo experiments.
  (c) Magnification of the dashed area in (b), showing timescales important for gates.
  The RB Ramsey data show a trend different from that predicted by the Ramsey fit. 
  }
  \label{fig:comparison}
\end{figure}

Quantum systems based on ion traps, spin qubits, and superconducting circuits are rapidly maturing, with individual operation fidelity at the levels required for fault-tolerant quantum computing \cite{Harty2014,Brown2011,Paik2011,Rigetti2012,Choi2014,Schindler2013,Awschalom2013,Itoh,Barends2014,Kelly2015}.
These systems are often limited by environmentally-induced phase noise, which can manifest as qubit frequency jitter.
Noise in the phase $\phi$ is characterized by variance $\braket{\phi^2 (\tau)}$, increasing linearly with time $\tau$ for white noise, and with higher power for correlated noise \cite{Martinis2003}.
Ramsey and spin echo experiments measure the decay of phase coherence for large magnitudes over long timescales; at much shorter timescales, which are relevant to quantum gates but still slower than the qubit frequency, dephasing errors are small and thus hard to measure, making physical mechanisms difficult to directly identify.
Here, we quantify phase noise by using RB to measure the decoherence of an identity gate versus its duration, providing an unprecedented metrological tool.

We use a superconducting quantum system based on the planar transmon qubit variant, the Xmon \cite{Barends2013}, cooled to 20 mK in a dilution refrigerator.
This qubit consists of a SQUID, which serves as a tunable non-linear inductor, and a large X-shaped shunt capacitor.
It is well-suited for characterizing phase noise as the qubit has long energy relaxation times, and the SQUID gives a controllable susceptibility to flux noise.
These qubits have frequencies that can be tuned to 6 GHz and below and have typical nonlinearities of $\eta/2\pi=-0.22$ GHz, and capacitive coupling strengths between qubits of $2g/2\pi=30$ MHz \cite{swapnote}.
Single qubit rotations are performed with microwave pulses and tuned using closed-loop optimization with RB \cite{Kelly2014}.
We use a dispersive readout scheme with capacitively coupled resonators at 6.6--6.8 GHz for state measurements \cite{Jeffrey2014}.
For details of the experimental setup and fabrication process, see \cite{Kelly2015}.

\section{RB Ramsey}

Figure~\ref{fig:comparison}a shows gate sequences for Ramsey and spin echo measurements, as well as their RB equivalents that we have called ``RB Ramsey'' and ``RB echo''.
The Ramsey experiment accumulates phase error from a single period $\tau$, whereas the RB Ramsey experiment accumulates phase error from $m$ applications of $\tau$, with $m$ typically of order 100.
In RB, gate error is measured directly by interleaving gates with random Clifford group operators, which depolarize errors by evenly sampling the Hilbert space, such that repeated gate applications add error incoherently \cite{Magesan2012}. 
Thus, RB Ramsey has a factor $m$ higher sensitivity than Ramsey when errors and times $\tau$ are small. The error of an idle gate, $r_{I(\tau)}$, is directly related to the variance of the phase noise by (see Appendix \ref{app:rbTheory})
\begin{equation} \label{eq:rbMain}
 r_{I(\tau)} = \frac{1}{6} \braket{\phi^2 (\tau)}.
\end{equation}
We infer and plot the equivalent Ramsey decay envelope visibility data $V$ (solid circles) with $V(\tau) = A\exp(-1/2 \braket{\phi^2(\tau)}) + B$ in Fig.~\ref{fig:comparison}b, with state preparation and measurement error (SPAM) parameters $A$ and $B$ extracted from the Ramsey fit as described in Appendix \ref{app:fitData}, and $\braket{\phi^2(\tau)}$ measured by RB Ramsey according to Eq. (\ref{eq:rbMain}).
We likewise show the equivalent spin echo decay envelope from RB echo data as solid squares.
The Ramsey and spin echo measurements over the same timescale are shown for comparison as open circles and open squares, respectively.
We label the length of a single qubit and two-qubit entangling gate \cite{Barends2014} to emphasize the relevant timescale.
The full Ramsey and spin echo measurements are shown on the typical linear scale, together with energy relaxation, in the inset of Fig.~\ref{fig:comparison}b.

As shown in Fig.~\ref{fig:comparison}b, the RB Ramsey and RB echo data are consistent with the Ramsey and spin echo measurements, respectively, at short to moderate time scales, while being much more precise.
Any structure to short-time dephasing is obscured in the Ramsey data, whereas the RB Ramsey data reveal a time dependence that we will show is consistent with telegraph noise. 
The use of RB greatly improves the precision of phase noise measurements; the uncertainty of the measured Ramsey visibility for $\tau < 300$ ns is reduced by an order of magnitude.
We note that the total time taken to perform the Ramsey and RB Ramsey measurements is approximately the same, and that precision would be increased for a higher-fidelity qubit by simply choosing larger $m$s.
Because of the imprecision of the Ramsey data at short time scales, the amount of noise present can only be inferred from the fit to the entire Ramsey dataset.
However, Fig.~\ref{fig:comparison}c shows that the phase noise measured by RB Ramsey can differ significantly from that expected by the Ramsey fit.
The trend in their difference indicates that there is behavior to the noise at short times that Ramsey measurements miss.
We examine this in Fig.~\ref{fig:rbRamsey}.

\section{Measuring Telegraph Noise}

\begin{figure}[b!]
  \centering
  \includegraphics[width=0.48\textwidth]{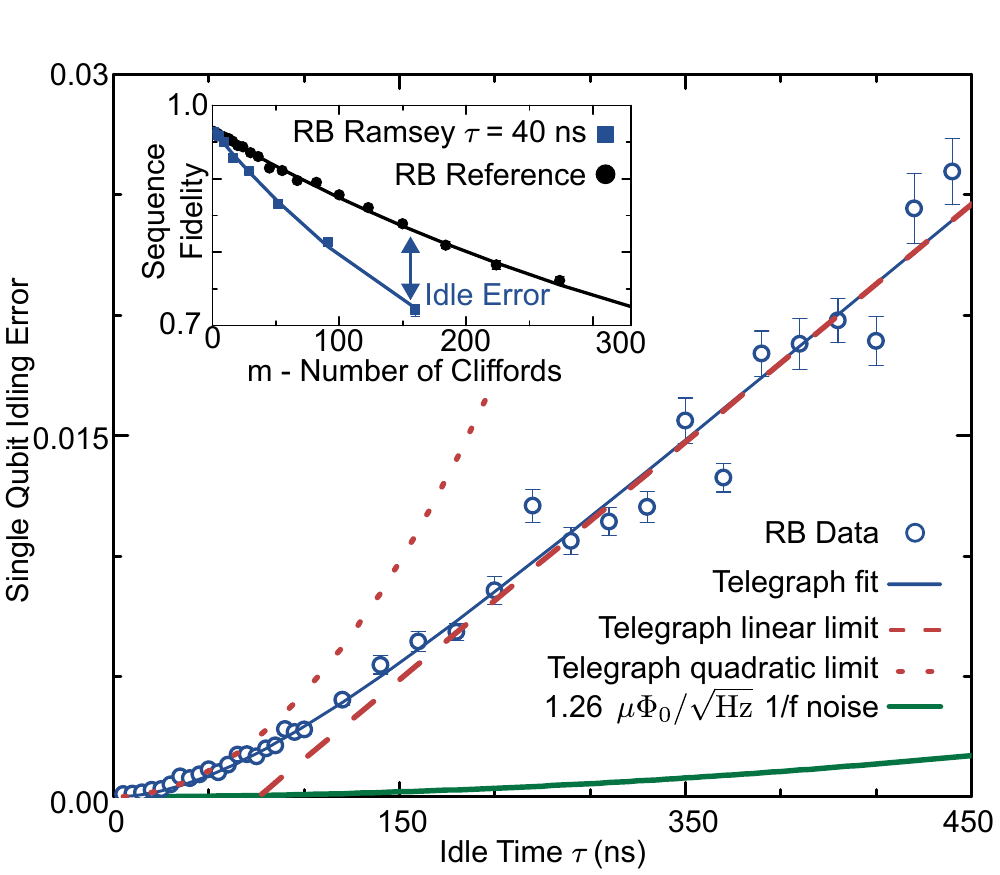}
  \caption{(color online) RB Ramsey measurement (circles) for  short
timescales; note that the small error from $T_1$ decay, which is $9\times 
10^{-4}$ at 450 ns, has been subtracted (see Appendix \ref{app:rbTheory}).
We fit to a telegraph noise model (Eq.~\ref{eq:telegraphFit}); the dotted
(dashed) lines give the short (long) time limit of the noise model. The
inferred but negligible contribution from $1/f$ noise  as measured for this 
qubit (see Appendix \ref{app:fluxNoise}) is shown  as a thick line. The inset shows the experiment 
used to extract the 40 ns data point.}
  \label{fig:rbRamsey}
\end{figure}

To identify the dominant noise mechanism, we examine the dependence of idle gate error on time and compare against different noise models in Fig.~\ref{fig:rbRamsey}.
Whereas in Fig.~\ref{fig:comparison} we infer an equivalent Ramsey envelope, here we plot the idle gate error directly, as measured by RB Ramsey (with small $T_1$ effects subtracted, see Appendix \ref{app:rbTheory}).
For short times, we see a non-linear increase of error with gate duration which transitions into a linear behavior for lengths above approximately 100 ns.
The inset shows the sequence fidelity \emph{vs.} number of Cliffords, with and without interleaved idles, used to extract the idle error for $\tau = 40$ ns.

While it has long been known that SQUIDs are susceptible to $1/f$ flux noise \cite{Paladino2002,Wellstood1987,Shnirman2005,Yoshihara2006,Bialczak2007,Sendelbach2008,Sank2012}, we find this a negligible contribution to gate error.
A system limited by $1/ f$ and white noise would see a linear increase in error at short, and quadratic increase at long times as the $1/f$ component begins to dominate.
The data exibit the opposite trend.
Moreover, the expected contribution to gate error from $1/f$ noise, as measured for this system below 1 Hz using the Ramsey Tomography Oscilloscope protocol (see \cite{Sank2012} and Appendix \ref{app:fluxNoise}), is significantly less than observed here (Fig.~\ref{fig:rbRamsey} thick solid line).

The trend observed in Fig.~\ref{fig:rbRamsey} is consistent with telegraph noise. For a random telegraph switching of the qubit frequency, the phase noise is given by
\begin{equation} \label{eq:telegraph}
\braket{\phi^2_\text{tel}(\tau)} = (2\pi\Delta f_{10})^2 T_{\text{sw}}
\left( \tau - T_\text{sw} \left[1-\exp\left(-\frac{\tau}{T_{\text{sw}}}\right) 
\right] \right),
\end{equation}
where $\Delta f_{10}$ is the effective switching amplitude of the qubit frequency and $T_{\text{sw}}$ is the switching timescale.
We make the simplifying assumption of symmetric telegraph noise as the measurement is unable to differentiate up and down switching rates, and note that while telegraph noise is not Gaussian, Eq. (\ref{eq:telegraph}) is still approximately correct for use in Ramsey and spin echo analyses (see Appendix \ref{app:rbTheory}).
In a more general case, the error rate for an idle of length $\tau$, $r_{I(\tau)}$, can be fit to a combination of error sources: white, long-time correlated, $1/f$, and telegraph phase noise, as well as $T_1$ decay,
\begin{align}
 r_{I(\tau)} = \frac{\tau}{3T_1} + \frac{1}{6} \Big(
	&\braket{\phi_\text{white}^2(\tau)} +	
	\braket{\phi_\text{corr}^2(\tau)} \nonumber \\
	&+ \braket{\phi_\text{1/f}^2(\tau)} +
	\braket{\phi_\text{tel}^2(\tau)} \Big) ,
     \label{eq:rbFull}
\end{align}
where the derivation for
$\braket{\phi_\text{white}^2(\tau)} = 2\tau/T_{\phi1}$, $\braket{\phi_\text{corr}^2(\tau)} = 2(\tau/T_{\phi2})^2$, 
and 
$\braket{\phi_\text{1/f}^2(\tau)}$
are given in Appendix \ref{app:phaseNoise}, and we assume correlated noise has a longer timescale than the experiment.
The data here are fitted to a noise model featuring only $T_1$ decay (measured independently) and telegraph noise,
\begin{equation} \label{eq:telegraphFit}
r_{I(\tau)} = \frac{\tau}{3T_1} + \frac{1}{6} \braket{\phi_\text{tel}^2(\tau)},
\end{equation}
indicating that $1/f$ and white noise do not dominate the error for this qubit.
We extract $T_{\text{sw}} = 84 \pm 14$ ns and $\Delta f_{10} = 479 \pm 30$ kHz from the fit.
The dotted (dashed) line shows this noise model in the short (long) time limit.
Perhaps surprisingly, this measurement directly shows that gates of duration 20 ns can achieve fidelity $>$ 0.999 in a system with characteristic Ramsey scale of $T_{\phi2} = 2.0 \, \mu$s (see Appendix \ref{app:fitData}).

Telegraph noise has been studied in superconducting circuits with a variety of methods.
Frequency fluctuations due to quasiparticle (QP) tunneling have been characterized by Rabi oscillations \cite{Bal2014} and repeated direct frequency measurement \cite{Riste2013}.
For our qubit, the calculated frequency splitting due to QP tunneling ranges from 1 Hz to 14 kHz (see Appendix \ref{app:chargeNoise}), well below the magnitude necessary to explain the data.
Photon shot noise in a coupled resonator has been shown to cause dephasing in both transmon \cite{Paik2011,Sears2012,Rigetti2012} and flux \cite{Stern2014} qubits.
In our case the magnitude of the telegraph noise decreases as the qubit--resonator frequency difference decreases, indicating that resonator photon noise induced dephasing is not the cause.
A more elusive telegraph-like noise has been measured by $T_{1\rho}$ Rabi spectroscopy in flux qubits \cite{Yan2013}, hypothesized to be due to two sets of coupled coherent two-level states.
This noise is similar in frequency to the noise measured here, with spectroscopic signatures at 1 and 20 MHz, compared to 1/84 ns = 11 MHz for this measurement.
However, it is much larger in magnitude, presenting as a ``dip'' (or ``plateau'') in spin echo measurements, which is known to happen in the presence of strong telegraph noise \cite{Galperin2006}, and seen in other systems \cite{Ithier2005,Stern2014,Riste2013}.
In our device, the telegraph noise is only dominant at short timescales, as any evidence of it in longer measurements like Ramsey and spin echo is masked by $1/f$ flux noise.

\section{Measuring Error from Coherent Qubit-Qubit Interactions}

\begin{figure}[b!]
  \centering
  \includegraphics[width=0.48\textwidth]{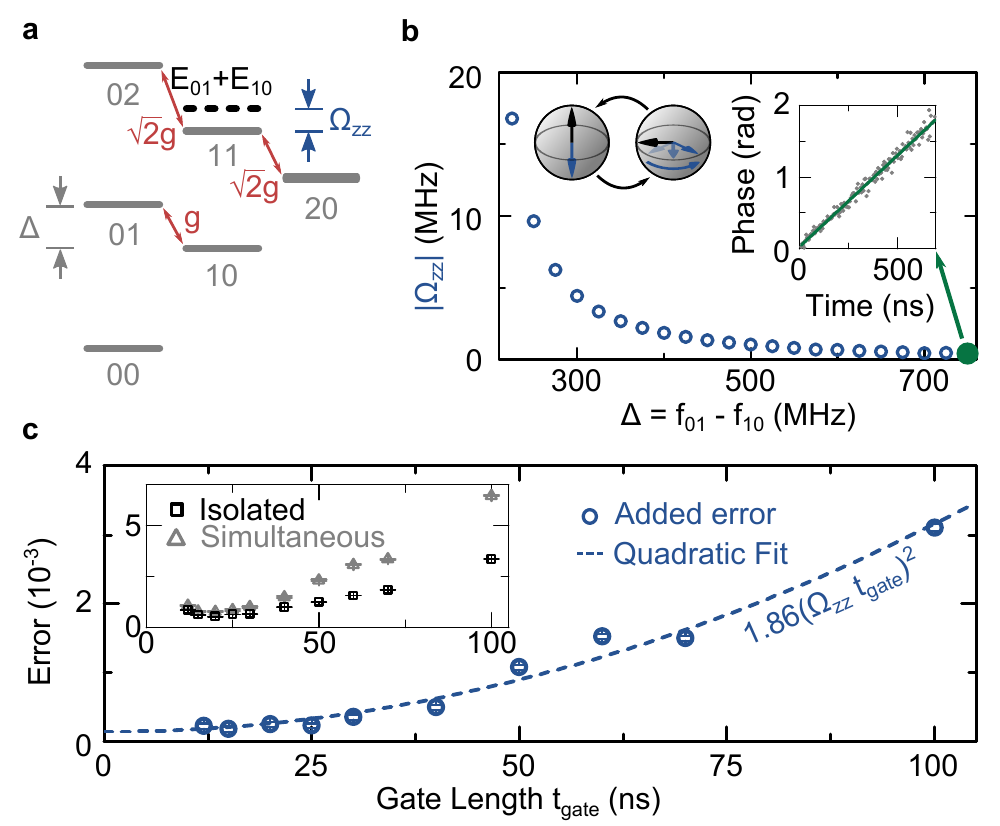}
  \caption{(color online)
  (a) Energy level diagram for two capacitively coupled
qubits with coupling strength $2g/2\pi = 30$ MHz, detuned by frequency $\Delta$.
The avoided level crossing between the $\ket{11}$ and $\ket{02}$/$\ket{20}$ 
states repels
the $\ket{11}$ frequency from the sum of $\ket{01}$ and $\ket{10}$ frequences by
the amount $\Omega_{ZZ}$.
  (b) This entangling interaction causes the phase of one qubit to
precess, conditional on the state of its neighbor (cartoon and inset). The
$\Omega_{ZZ}$ interaction decreases with $\Delta$, to a level of
$\Omega_{ZZ}/2\pi = 0.4$ MHz at $\Delta/2\pi = 750$ MHz.
  (c) RB data isolating the $\Omega_{ZZ}$
interaction. Gate error is measured \emph{vs.} gate duration for a single 
qubit and when
qubits are operated simultaneously (inset). The difference (main figure) 
measures the
error contribution from the $\Omega_{ZZ}$
interaction, and is fit to $1.86 (\Omega_{ZZ} t_\text{gate} / 2\pi)^2 + 
1.4\cdot 10^{-4}$. }
  \label{fig:omegaZZ}
\end{figure}

We now apply RB to coherent errors arising from unwanted qubit-qubit interactions, which can also contribute to dephasing \cite{DiCarlo2009}.
In Fig.~\ref{fig:omegaZZ}, we explore these effects in our system.
Figure~\ref{fig:omegaZZ}a shows an energy level diagram for capacitively coupled qubits, where the fundamental entangling rate $\Omega_{ZZ}$ \cite{Galiautdinov2012} arises from an avoided level crossing between the $\ket{11}$ state and the $\ket{02}$ and $\ket{20}$ states.
This interaction manifests as a state-dependent frequency shift, falling off with detuning $\Delta$, as measured in Fig.~\ref{fig:omegaZZ}b.
We note that for a qubit coupled to a resonator, $\Omega_{ZZ}$ is equivalent to the dispersive shift \cite{Blais2004} $2\chi$ as defined in \cite{Koch2007}.
The inability to turn this interaction off completely results in additional errors when operating qubits simultaneously.
Figure~\ref{fig:omegaZZ}c shows average gate error \emph{vs.} duration, when a qubit is operated in isolation or simultaneously with a coupled qubit ($\Omega_{ZZ}/2\pi$ = 0.4 MHz).
Error for single qubit or simultaneous operation is inferred from the RB reference error per Clifford, divided by the average of 1.875 physical gates per Clifford \cite{Barends2014}.
The difference between isolated and simultaneous operation gives the added error from the $\Omega_{ZZ}$ interaction, which is fit to a quadratic.

This interaction is correlated, and therefore the errors are quadratic with gate duration; specifically, the error per gate due to the $\Omega_{ZZ}$ interaction between two qubits simultaneously undergoing RB is
\begin{equation} \label{eq:omegaZZtheory}
 E = \frac{\pi^2}{6} \left(\frac{\Omega_{ZZ}}{2\pi} t_\text{gate} \right)^2,
\end{equation}
where $\Omega_{ZZ}/2\pi$ is the interaction magnitude and $t_\text{gate}$ is the RB gate duration (see Appendix \ref{app:omegaZZ}).
The fit to the data has a quadratic coefficient of $1.86 \pm 0.1$, while $\pi^2/6 \approx 1.64$.
Here, the careful application of RB both distinguishes these errors at the $1\cdot10^{-4}$ level, and indicates that short gates are effective in suppressing them.

\section{Measuring Different Gate Implementations}

\begin{figure}[b!]
  \centering
  \includegraphics[width=0.48\textwidth]{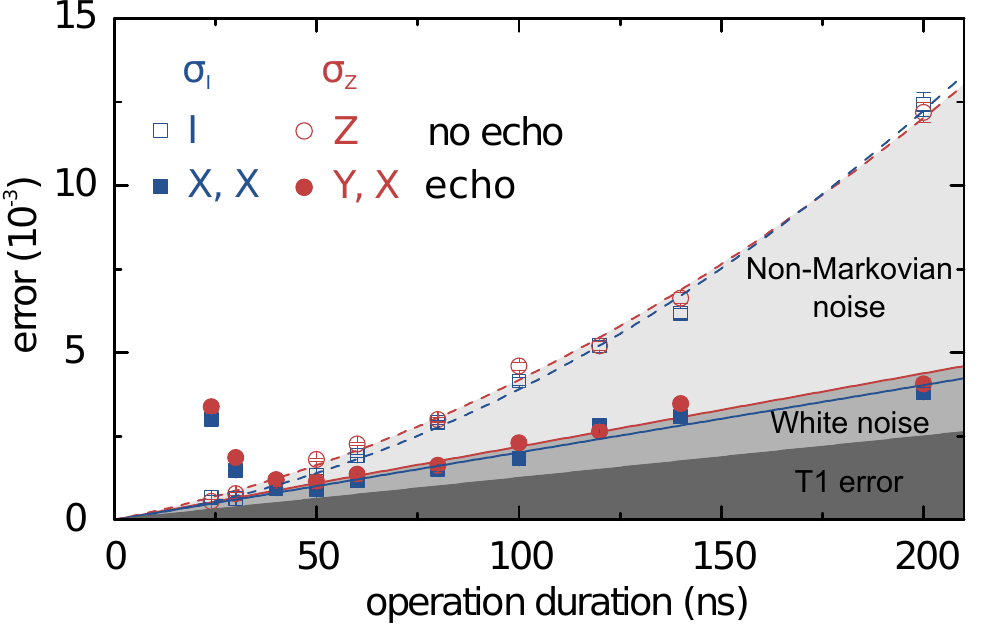}
  \caption{(color online)
  Operation error of $\sigma_I$ and $\sigma_Z$, implemented with (closed symbols) and without (open symbols) echoing, as measured with interleaved RB. The data are fitted to a linear and quadratic form, representing uncorrelated and correlated noise.
  The dark gray region indicates error attributed to $T_1$, the medium gray region uncorrelated noise, and the light gray region non-Markovian (e.g. telegraph) noise.
  Note that the $I$ data are RB Ramsey data, the same as Fig.~\ref{fig:rbRamsey}.
  }
  \label{fig:gates}
\end{figure}

We now examine the gate fidelity for a variety of gates in the presence of the non-Markovian noise we have measured.
Figure ~\ref{fig:gates} shows gate fidelity \emph{vs.} gate length for two implementations each of two different gates: for $\sigma_I$, an idle and two microwave pulses ($X$, $X$), and for $\sigma_Z$, a frequency detuning pulse and two microwave pulses ($Y$, $X$).
The errors of these operations \emph{vs.} duration are determined with interleaved RB.
In agreement with previous measurements, we find that the error of operations without $X$ or $Y$ pulses (open symbols) follow a quadratic-like dependence with gate duration at these timescales.
Using $X$ or $Y$ pulses (closed symbols), we observe a linear-like dependence at longer durations, indicating that the correlated phase noise has been suppressed.
Below 40 ns, we find an increased error which we attribute to the population of higher levels due to spectral leakage \cite{Lucero2010}.
The solid (dashed) lines are linear (linear and quadratic) fits to the data. For full details of the fits, see Appendix \ref{app:fitData}.

Using the functional forms of the different error types given in Eq.~(\ref{eq:rbFull}), we can determine an error budget for our operations.
For a typical entangling gate duration of 40 ns, $T_1$ contributes an error of $5\cdot 10^{-4}$, and telegraph noise an error of $5\cdot 10^{-4}$.
With echoing pulses, the total error is $8\cdot 10^{-4}$, indicating that the added echoing pulses are either not completely suppressing the phase noise or are contributing error of their own.
Using a combination of RB Ramsey and RB echo, we have determined the relative contribution of different noise sources to operational error, and we can also immediately see that either short gates, or long gates with intrinsic echoing, are effective at remedying non-Markovian noise, and by how much.

\section{Summary}

RB Ramsey provides a direct measurement of phase noise in the regime most relevant to quantum gates.
While previous noise spectroscopy has relied on accumulating noise over longer timescales while filtering out low-frequency noise with additional pulses, our technique directly measures small amounts of noise with repeated incoherent additions.
It does not require extensive calibration, and is also robust against state preparation and measurement error.
As a gate-based measurement, it is useful in a variety of situations: measuring noise due to the environment as RB Ramsey, measuring filtered environmental noise as RB echo, and measuring dephasing induced by coherent qubit-qubit interactions.
As the measurement output is gate fidelity, it is also immediately applicable as a tool to determine the highest-fidelity implementation of different quantum gates in the presence of noise.
We show here that RB Ramsey is the metrological tool best suited for measuring noise in high-fidelity qubits.

We have taken RB, a protocol for determining the fidelity of gates, and applied it as a metrological tool for identifying noise processes.
Applied to a superconducting qubit system, we have found a telegraph noise mechanism in a regime inaccessible to previous measurements, accurately characterized dephasing caused by coherent qubit-qubit interactions, and determined the highest-fidelity implementation of different quantum gates.
Our results demonstrate that RB Ramsey is capable of measuring small noise processes at short timescales that are directly relevant to gate fidelity, and show that understanding this non-Markovian phase noise can be lead to its effective suppression through short gates and echoing.

\begin{acknowledgments}
We thank A. Veitia for helpful discussions on the theory of noise and RB.
This work was supported by the Office of the Director of National Intelligence (ODNI), Intelligence Advanced Research Projects Activity (IARPA), through the Army Research Office grants W911NF-09-1-0375 and W911NF-10-1-0334.
All statements of fact, opinion or conclusions contained herein are those of the authors and should not be construed as representing the official views or policies of IARPA, the ODNI, or the U.S. Government.
Devices were made at the UC Santa Barbara Nanofabrication Facility, a part of the NSF-funded National Nanotechnology Infrastructure Network, and at the NanoStructures Cleanroom Facility.
P. J. J. O'Malley, J. Kelly, and R. Barends contributed equally to this work. 
\end{acknowledgments}

\appendix

\section{Theoretical relation of RB error to $\braket{\phi^2}$}
\label{app:rbTheory}
In order to determine the effect of various dephasing mechanisms on an RB sequence, we first consider the following simplified model: a single qubit begins in $\ket{\psi_0} = \ket{0}$, then a randomly chosen perfect Clifford rotation $C_1$ is applied, and then a phase $\phi_{g,n}$ is accumulated by application of a Z rotation to simulate phase noise.
The random Clifford and noise gate pair are repeated $N$ times, after which the single Clifford $C_r$ that is the inverse of all the previous  Cliffords is applied to rotate back to (nearly) $\ket{0}$ and we measure the probability of error, $P_{err} = \abs{\braket{1|\psi_N}}^2$.

The value of $\phi_{g,n}$ depends on the dephasing model employed. For example, for static dephasing (e.g., a frequency offset), it is constant: $\phi_{g,n} = \phi_{g, st}$. For white noise,  $\phi_{g,n}$ is randomly sampled from a symmetric Gaussian distribution. In general, $\phi_{g,n}$ is arbitrary, but we assume $|\phi_{g,n}|\ll 1$. The average square of $\phi_{g,n}$ is denoted $\braket{\phi_g^2}$.

We now consider the ``error angle'', $\Delta\phi$, the angular separation of
$\ket{\psi_N}$ from $\ket{0}$ in the Bloch sphere picture of a single qubit,
noting that $P_{err} = \braket{(\Delta\phi/2)^2}$, assuming $|\Delta\phi|\ll 1$. Because $|\phi_{g,n}| \ll 1$ and $N$ is not too large,
after each rotation $\ket{\psi}$ is close to one of the six axes ($\pm X, \pm Y, \pm Z$), and the angular distance from the axis is $\Delta\phi$. There is a 1/3
chance that the qubit is near the pole (i.e. Z axis) and then the rotation
$\phi_{g,n}$ does not change $\Delta\phi$, while with 2/3 probability the qubit is near the equator and $\Delta\phi$ is changed.

For any dephasing model, it is straightforward to see that the evolution
of $\Delta\phi$ is essentially a random walk in two dimensions, and that
\begin{equation} \label{eq:errorAngle}
  \braket{(\Delta\phi)^2} = \frac{2}{3}N\braket{\phi_g^2},
\end{equation}
assuming $N\braket{\phi_g^2} \ll 1$. The RB error is then
\begin{equation} \label{eq:rbError}
 P_{err} = \braket{(\Delta\phi/2)^2}=\frac{1}{6}N\braket{\phi_g^2}.
\end{equation}

It might be expected that in the static dephasing case---when there are correlated phase contributions---there can be some sort of echoing effect; for example, if a Clifford takes $\ket{\psi}$ to the +Y axis and it is rotated by $\phi_{g,st}$, then if the next Clifford is an X rotation, putting $\ket{\psi}$ near the -Y axis, the following rotation also by $\phi_{g,st}$ will cancel the previous noise rotation.
However, when the full set of Clifford rotations is used, there are four rotations that take $\ket{\psi}$ near the -Y axis, and each orients the previous $\Delta\phi$ in a different direction relative to the axis, resulting in equal probability of canceling the previous rotation, doubling it, or moving in one of the two perpendicular directions.
The noise accumulated between rotations is therefore uncorrelated with previous or future noise; the Clifford set is error depolarizing.
Therefore, Eqs.~(\ref{eq:errorAngle}) and (\ref{eq:rbError}) hold regardless of the noise model.

This simplified model has been confirmed with simulation, for both a static and
an uncorrelated noise model with $\phi_{g,n}=\pm \phi_{g}$.

This implies that RB is an effective way to measure dephasing, if
the sequence error occuring between the gates is attributable to dephasing.
This can be done easily by comparing the sequence fidelity of an RB sequence
with interleaved idling time to that of a reference RB sequence, effectively
subtracting out errors due to the Clifford gates themselves--in other words,
measuring the fidelity of an idle using interleaved RB, as in
\cite{Barends2014}. We can therefore measure the dephasing that takes place
during an idle, and by varying the length $\tau$ of an idle, measure
dephasing as a function of time, $\braket{\phi^2(\tau)}$ (for brevity we removed the subscript $g$). With $r_{I(\tau)}$
being the error rate (i.e. error per gate) of an idle, we thus arrive at
Eq.~(\ref{eq:rbMain}):
\begin{equation} \label{eq:rbErrorDerivation}
 P_{err}/N = r_{I(\tau)} = \frac{1}{6}\braket{\phi^2(\tau)}.
\end{equation}

For completeness, we also mention here the effect of energy relaxation ($T_1$
decay) on the fidelity of RB sequences. After each Clifford, the qubit state $\ket{\psi}$ is near the equator of the Bloch sphere with probability 2/3. In this case the probability of the energy relaxation event is $\tau/2T_1$ (we assume $\tau \ll T_1$); such an event moves $\ket{\psi}$ by approximately the angle $\pi/2$ on the Bloch sphere, thus leading to the error probability 1/2 at the end of the RB sequence. The corresponding contribution to the RB error per gate is $(2/3)\times (\tau/2T_1)\times (1/2)=\tau/6T_1$. With probability 1/6 the qubit state after a Clifford is close to the North pole (state $\ket{0}$); then there is no energy relaxation. Finally, with probability 1/6 the qubit state is close to the South pole  $\ket{1}$; then the probability of the energy relaxation event is $\tau/T_1$, which moves the state by approximately the angle $\pi$, thus almost certainly leading to the RB error. The corresponding contribution to the RB error per gate is $(1/6)\times (\tau/T_1)\times 1=\tau/6T_1$. Adding together the two contributions, we arrive at
\begin{equation} \label{eq:rbT1}
 P_{err}/N = \frac{\tau}{3 T_1}.
\end{equation}
Since $T_1$ can be measured independently, the effects of $T_1$
decay can be calculated and subtracted from the results obtained
with RB, much as it can be subtracted from Ramsey visibility decays
as well. In our experiment $T_1$ is relatively large, and therefore
this correction is small.

\section{Types of Phase Noise}
\label{app:phaseNoise}

We now discuss the form of $\braket{\phi^2(\tau)}$ for
different sources of noise. For completeness, we also show the similar characteristic,  $\braket{\tilde\phi^2(\tau)}$, for the echo sequence of duration $\tau$ (with $\pi$ pulse at $\tau/2$). Most of results discussed here were presented earlier, e.g., in Refs.\ \cite{Cottet2002}\cite{Martinis2003}\cite{Yoshihara2006}.

The average values $\braket{\phi^2(\tau)}$ and $\braket{\tilde\phi^2(\tau)}$ for the idle and echo sequence, respectively, can be calculated via the spectral density $S(\omega )$ of the qubit frequency fluctuation,
    \begin{eqnarray}
&&    \braket{\phi^2(\tau)} = \tau^2 \int_0^\infty S(\omega) \left(\frac{\sin (\omega\tau/2)}{\omega\tau/2}\right)^2 \frac{d\omega}{2\pi},
    \label{eq:phi2-S}\\
&&   \braket{\tilde\phi^2(\tau)} = \tau^2 \int_0^\infty S(\omega) \frac{\sin^4 (\omega\tau/4)}{(\omega\tau/4)^2} \frac{d\omega}{2\pi},
    \label{eq:phi-tilde-S}\end{eqnarray}
where $S(\omega)$ is single-sided and the average frequency fluctuation is assumed to be zero.

For the white noise  with a flat spectral density, $S(\omega) = S_0$, we find
\begin{equation} \label{eq:whiteNoise}
 \braket{\phi_\text{white}^2(\tau)} =  \braket{\tilde\phi_\text{white}^2(\tau)} = \frac{S_0}{2} \tau = 2 \frac{\tau}{T_{\phi 1}},
\end{equation}
where $T_{\phi 1}=4/S_0$ is the dephasing time due to white noise. Note that the factor of 2 in the last expression cancels when the corresponding visibility of a Ramsey or echo sequence,  $\exp (-\tau/T_{\phi 1})$, is calculated.

For noise that is correlated over very long times (very slowly fluctuating qubit frequency), $S(\omega)=4\pi\sigma_{qb}^2 \delta (\omega)$, where $\sigma_{qb}$ is the standard deviation of the qubit frequency $2\pi f_{10}$. In this case
\begin{equation} \label{eq:correlatedNoise}
 \braket{\phi_\text{corr}^2(\tau)} = \sigma^2_{qb} \tau^2 =
    2 \left(\frac{\tau}{T_{\phi2}}\right)^2, \,\,\,  \braket{\tilde \phi_\text{corr}^2(\tau)}=0,
\end{equation}
where $T_{\phi2}=\sqrt{2}/\sigma_{qb}$ is the Ramsey dephasing timescale due to such correlated noise. Obviously, in this case there is no dephasing in the echo sequence visibility.

For $1/f$ noise let us use
$S(\omega) = \displaystyle \frac{ S_{1/f}}{\omega/2\pi}$,
then \cite{Martinis2003,Yoshihara2006}
\begin{eqnarray} \label{eq:1fNoise}
 && \braket{\phi_{1/f}^2(\tau)} = S_{1/f}\, \tau^2 \ln\frac{0.4007}{f_c \tau},
  \\
  && \braket{\tilde\phi_{1/f}^2(\tau)} = S_{1/f}\, \tau^2 \ln 2,
\label{eq:1fNoise-tilde}\end{eqnarray}
where $f_c=\omega_c/2\pi$ is the low-frequency cutoff of the $1/f$ noise (e.g., 
the inverse of the total duration of the experiment), which is introduced as 
the lower limit of integration in Eq.~(\ref{eq:phi2-S}). Note that in 
Eq.~(\ref{eq:1fNoise}) we assumed $f_c \tau \alt 0.2$. As the log part in 
Eq.~(\ref{eq:1fNoise}) varies slowly, typically it is ignored and $1/f$ noise 
for $\braket{\phi^2(\tau)}$ is treated with Eq.~(\ref{eq:correlatedNoise}). 
Note that the factors in Eq.\ (\ref{eq:1fNoise}) and (\ref{eq:1fNoise-tilde}) 
are different, resulting in different effective dephasing times $T_{\phi2}$ for 
the Ramsey and echo sequences.

Finally, let us consider a telegraph noise, for which the qubit frequency $2\pi f_{10}$ switches between two
values separated by $\Delta\omega_{qb}$, with up (down) switching rate of
$\Gamma_\uparrow$ ($\Gamma_\downarrow$). In this case
  \begin{equation}
S(\omega)=\frac{4(\Delta\omega_{qb})^2 \Gamma_\uparrow\Gamma_\downarrow}{\Gamma_{\Sigma} (\omega^2+\Gamma_\Sigma^2)}, \,\,\, \Gamma_\Sigma = \Gamma_\uparrow +\Gamma_\downarrow,
    \end{equation}
so using Eqs.\ (\ref{eq:phi2-S}) and (\ref{eq:phi-tilde-S}) we obtain
\begin{equation} \label{eq:telegraphNoisePreliminary}
  \braket{\phi_\text{tel}^2(\tau)} = 2 \frac{(\Delta\omega_{qb})^2}{\Gamma_\Sigma}
    \frac{\Gamma_\uparrow \Gamma_\downarrow}{\Gamma_\Sigma^2}
    \left(\tau - \frac{1-e^{-\Gamma_\Sigma \tau}}{\Gamma_\Sigma} \right),
    \end{equation}
    \begin{equation}
     \braket{\tilde \phi_\text{tel}^2(\tau)} = 2 \frac{(\Delta\omega_{qb})^2}{\Gamma_\Sigma}
    \frac{\Gamma_\uparrow \Gamma_\downarrow}{\Gamma_\Sigma^2}
    \left(\tau - \frac{3+e^{-\Gamma_\Sigma \tau}-4e^{-\Gamma_\Sigma \tau/2}}{\Gamma_\Sigma} \right) .
\end{equation}
Note that at short time, $\tau \ll \Gamma_\Sigma^{-1}$, the effect of the telegraph noise is similar to the effect of the correlated noise with $T_{\phi2}=\sqrt{2}\, \Gamma_\Sigma/(\sqrt{\Gamma_\uparrow\Gamma_\downarrow} \, \Delta\omega_{qb})$, while at long time, $\tau \gg \Gamma_\Sigma^{-1}$ it is similar to the effect of white noise with $T_{\phi1}=\Gamma_\Sigma^3/[\Gamma_\uparrow\Gamma_\downarrow(\Delta\omega_{qb})^2]$.

Defining the effective switching amplitude as $2\pi\Delta f_{10}=2 \Delta\omega_{qb}\sqrt{\Gamma_\uparrow\Gamma_\downarrow}/\Gamma_\Sigma$ and introducing notation  $T_\text{sw} = 1/\Gamma_\Sigma$, we can rewrite Eq.\ (\ref{eq:telegraphNoisePreliminary}) as
\begin{equation} \label{eq:telegraphNoise}
 \braket{\phi^2_\text{tel}(\tau)} = (2\pi\Delta f_{10})^2 T_{\text{sw}}
[ \tau - T_\text{sw} (1-e^{-\tau/T_{\text{sw}}})], \quad
\end{equation}
which is Eq.~(\ref{eq:telegraph}).
In the case where $\Gamma_\uparrow = \Gamma_\downarrow$, as we have assumed, $2 \pi \Delta f_{10}$ provides a lower bound on $\Delta \omega_{qb}$.
Note that the telegraph noise in not Gaussian.
Therefore, while the obtained equations can be used to find the RB error per gate, they cannot, strictly speaking, be used to find the visibility of the standard Ramsey and echo sequences.
Nevertheless, they can be used approximately if $|\Delta\omega_{qb}|/\min (\Gamma_\uparrow,\Gamma_\downarrow)\ll 1$, because at short time the accumulated phase shift is small and the assumption of Gaussianity is not needed, while at longer time, when the phase becomes comparable to 1, the probability distribution for the phase becomes Gaussian due to a large number of switching events.

\section{$T_1$, Ramsey, and spin echo fits}
\label{app:fitData}
The $T_1$ data are fit to a simple exponential, $P_1(t) = A\exp(t/T_1)+B$, and 
we find $T_1 = 26.7 \, \mu$s. The Ramsey and spin scho envelopes are each fit 
to a noise model that includes white and correlated components,
\begin{equation} \label{eq:ramseyFit}
 V(t) = A \exp \left[ -t/T_{\phi1} - (t/T_{\phi2})^2 \right] + B,
\end{equation}
where $V(t)$ is the Ramsey/echo visibility, $t$ is the length of the idle as
shown in Fig.~\ref{fig:comparison}, $T_{\phi1}$ is the white noise dephasing
timescale, $T_{\phi2}$ is the correlated noise dephasing timescale, and $A$ and
$B$ are the result of state preparation and measurement errors. The fit parameters are
given below. Note that each of the fits includes the full range of data, from $0 < t < 5.0
\, \mu$s for Ramsey and $0 < t < 12.0 \, \mu$s for Echo.

\begin{center}
\begin{tabular}{c|c|c|c|c}
  Sequence  & $T_{\phi1}$    & $T_{\phi2}$    & A & B \\
            & ($\mu$s)       & ($\mu$s)       &   &   \\ \hline
  Ramsey    & 6.8            & 2.8            & 0.88 & 0.015 \\ \hline
  Spin Echo & 15.1           & 7.5            & 0.88 & 0.021 \\
\end{tabular}
\end{center}

\section{Flux noise}
\label{app:fluxNoise}

\begin{figure}[h!]
  \centering
  \includegraphics[width=0.48\textwidth]{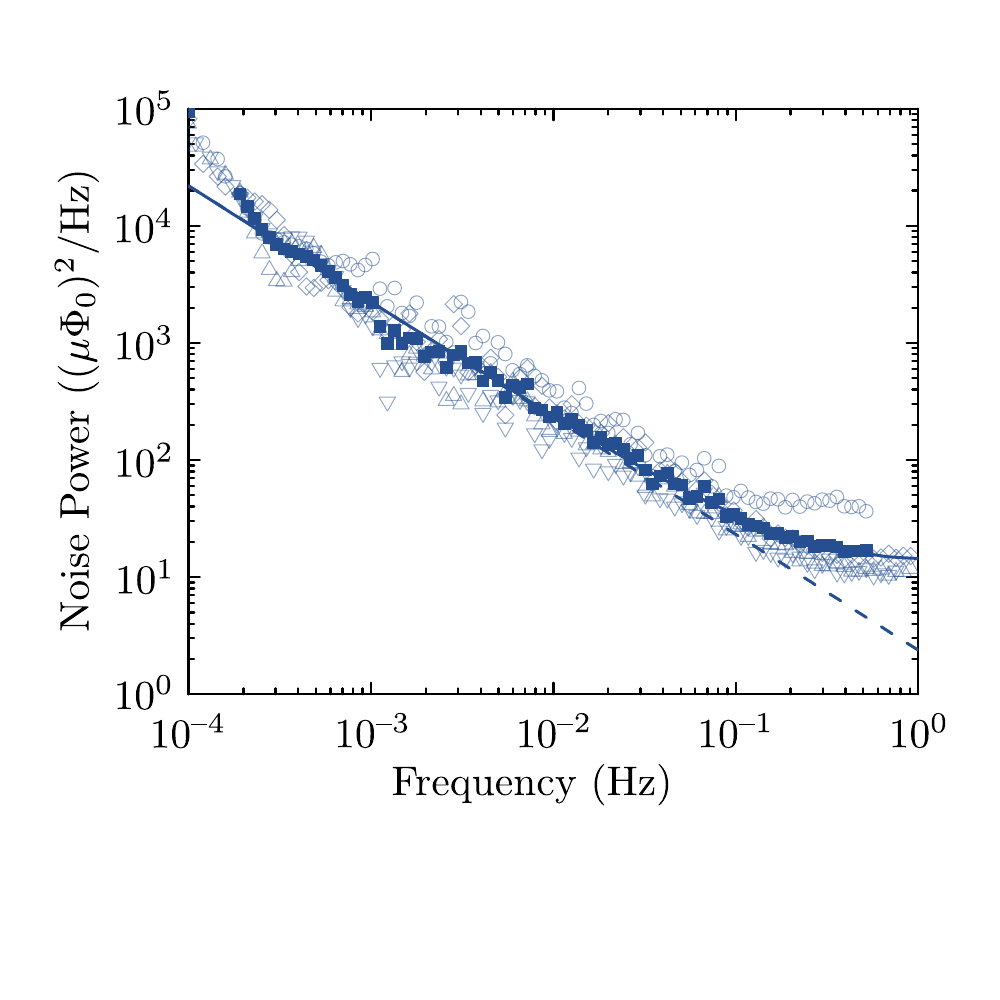}
  \caption{Flux noise as measured with RTO \cite{Sank2012}}
  \label{fig:fluxNoise}
\end{figure}

Flux noise on this device, plotted in Fig.~\ref{fig:fluxNoise}, has been 
measured over the frequency range $10^{-4} < f < 1$ Hz, using the Ramsey 
Tomography Oscilloscope (RTO) protocol of repeated frequency measurements 
as described in \cite{Sank2012}. Four measurements were made on 
this device (open markers), at three different operating points, and then each 
measurement was binned in log-space, and the binned measurements averaged 
together (closed squares). This average is fit (solid line) to an aliased $1/f$ 
and white noise model, given by
\begin{equation} \label{eq:fluxNoise}
 S_\phi(f) = S^*_\phi / f^\alpha + S^*_\phi/(2 f_n - f)^\alpha +
S_{\text{white}},
\end{equation}
where $S_\phi(f)$ is the flux noise power, expressed in $(\mu\Phi_0)^2/$Hz, $f$ 
is the noise frequency, $\alpha$ is the slope of the noise (1 for pure $1/f$
noise), $S^*_\phi$ is the flux noise power at 1 Hz, $f_n=1$ Hz is the Nyquist
frequency of the measurement, and $S_\text{white}$ is the white noise floor.
From the fit we extract $S^*_\phi = 2.4 \, (\mu\Phi_0)^2$, $\alpha = 0.99$, and
$S_\text{white} = 9.7 \, (\mu\Phi_0)^2/$Hz. We attribute the white noise to
state preparation and measurement error. The dashed line shows the $1/f$ fit
extended to 1 Hz, where the value of the y-intercept is $S^*_{\phi}$.

To plot the inferred flux noise contribution in Fig.~\ref{fig:rbRamsey} and Fig.~\ref{fig:fluxSensitivities} below, we use Eq.~(\ref{eq:1fNoise}), with $S_{1/f}=\partial f / \partial \phi \cdot S^*_\phi$ taken from the measurements above, and $f_c = 10$ min, the length of the experiment.
The value of the log factor of Eq.~(\ref{eq:1fNoise}) varies from 13 to 7 for $1 < \tau < 450$ ns.

This analysis assumes that the low frequency flux noise measured here can be extrapolated to high frequencies.
In Fig.~\ref{fig:rbRamsey}, however, we see that this calculation underestimates the amount of high frequency noise, and furthermore, that the noise is telegraph in nature, not $1/f$.

\section{RB Ramsey across the qubit spectrum}
\label{app:rbSpectrum}

\begin{figure}
  \centering
  \includegraphics[width=0.48\textwidth]{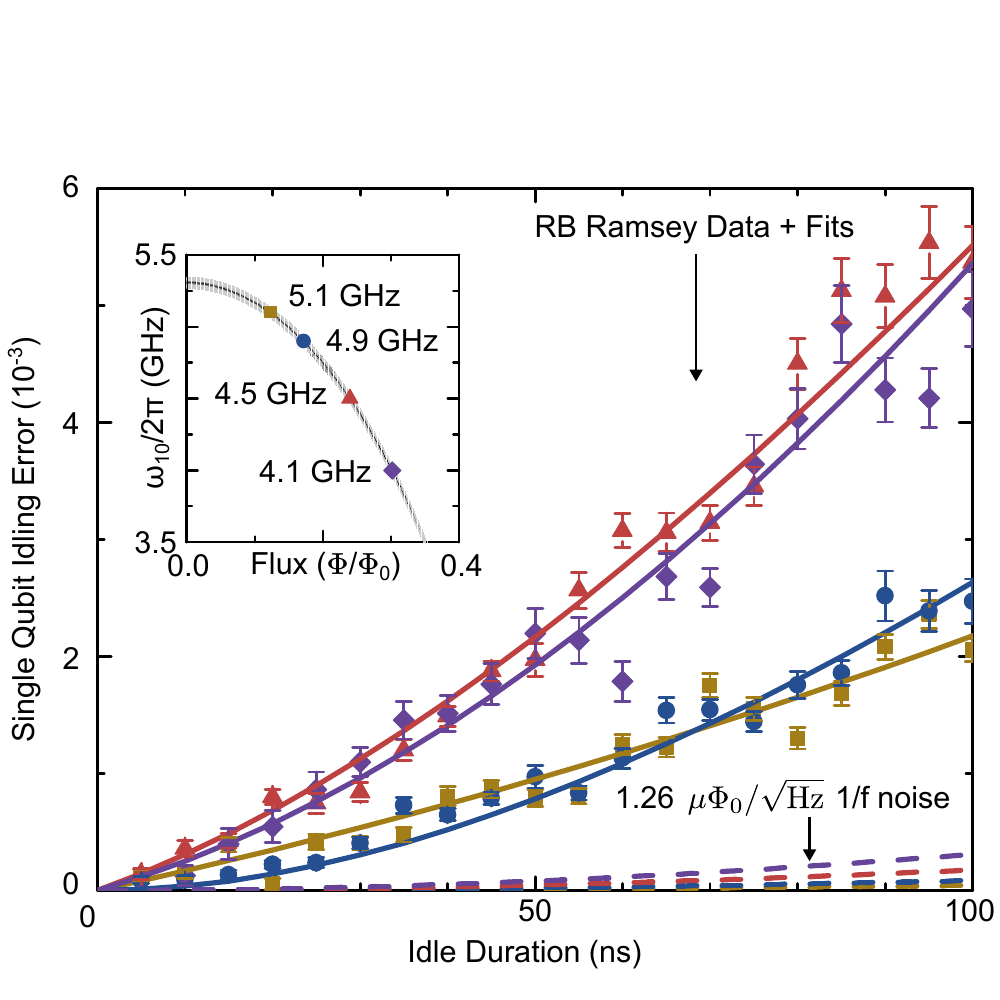}
  \caption{(color online) RB Ramsey idling error \emph{vs.} duration, for various
frequencies; $T_1$ effects have been subtracted according to 
Eq.~(\ref{eq:rbT1}). The dashed lines denote the inferred contribution from 
$1/f$ flux noise at the four different operating points. The inset
shows frequency spectroscopy \emph{vs.} applied flux, following the expected dependence
\cite{Koch2007}; the four operating points are shown.}
  \label{fig:fluxSensitivities}
\end{figure}

Figure~\ref{fig:fluxSensitivities} shows RB Ramsey measurements at three 
additional qubit frequencies; the data for the 4.9 GHz operating point are the 
same as in Fig.~\ref{fig:rbRamsey}. The inset shows the frequency-flux 
relation for this qubit, with the four operating points denoted by symbols; 
$df/d\phi$ changes by a factor of 2.7 between the operating points, to explore 
different susceptibilities to flux noise. The remaining three datasets are fit 
to a noise model incorporating telegraph and white noise; that is,
\begin{equation} \label{eq:fig3fit}
 r_{I(\tau)} = \frac{1}{6} \left( \braket{\phi_\text{tel}^2(\tau)} +
\braket{\phi_\text{white}^2(\tau)} \right)
\end{equation}
[see Eqs.~(\ref{eq:whiteNoise}) and (\ref{eq:telegraphNoise})]. We show the fit parameters 
here.
\begin{center}
 \begin{tabular}{c|c|c|c|c|c}

   $f_{10}$ & $df/d\Phi$ & $T_1$ &  $T_{\phi1}$ &
   $T_{\text{sw}}$ & $\Delta f_{10}$  \\
  (GHz) & (GHz/$\Phi_0$) & ($\mu$s) &
  ($\mu$s) & (ns) & (kHz)  \\ \hline
  5.1 & 3.39 & 30.6 & 20.6 & 182000 & 184 \\ \hline
  4.9 & 4.81 & 26.7 & - & 84 & 479 \\ \hline
  4.5 & 6.95 & 31.3 & 12.4 & 98 & 484 \\ \hline
  4.0 & 9.23 & 36.2 & 15.5 & 263 & 469

 \end{tabular}
\end{center}
We note that at the highest qubit frequency, the large $T_{\text{sw}}$ indicates that the telegraph noise model is not needed here and can be replaced by the
correlated noise model with $T_{\phi2} = \sqrt{2}/[2\pi \Delta f_{10}] = 1.2 \, \mu$s. The 
Ramsey data
for this frequency, fit to Eq.~(\ref{eq:ramseyFit}), give $T_{\phi1} = 10.7 \, 
\mu$s and
$T_{\phi2} = 3.6 \, \mu$s, which indicates that even though the telegraph dephasing source is
not present at this operating point, the dephasing magnitude measured by Ramsey still does not 
match that found with RB.

Despite tuning the flux $\Phi/\Phi_0$ over most of its range, we find that 
$1/f$ noise does not contribute appreciably to gate errors. For typical gates 
of length 20ns, idle fidelities greater than 0.999 are seen over the frequency 
range, demonstrating that tunable qubits can achieve high fidelity even when 
biased significantly away from the flux-insensitive operating point.

\section{Charge Noise}
\label{app:chargeNoise}

To calculate the expected frequency fluctuation due to charge noise, we use
Eq.~(2.5) from \cite{Koch2007}
\begin{equation} \label{eq:chargeNoise}
 \epsilon_m \simeq (-1)^m E_C \frac{2^{4m+5}}{m!} \sqrt{\frac{2}{\pi}}
 \left(\frac{E_J}{2E_C}\right)^{\frac{m}{2}+\frac{3}{4}} e^{-\sqrt{8E_J/E_C}} ,
\end{equation}
where $\epsilon_m$ is the charge dispersion for energy level $m$, and $E_J$ and
$E_C$ are the Josephson energy and charging energy, respectively, of the
qubit. Note that we can also write
$E_J/E_C \approx \left( \omega_{01}/\eta - 1 \right)^2 / 8$ (following from Eq.~
(2.11)),
where $\omega_{01}/2\pi$ is the qubit frequency and the qubit
anharmonicity $f_{12} - f_{01} = \eta/2\pi = -215$ MHz. We then calculate 
$\epsilon_1$ for the two
ends of the qubit spectrum; we find $\epsilon_1(\omega_{01}/2\pi = 6
\textrm{ GHz}) = 3.6$ Hz and $\epsilon_1(\omega_{01}/2\pi = 4 \textrm{ GHz}) =
14.4$ kHz, both of which are far below the measured charge noise fluctuation
frequency of $\approx 500$ kHz. We also note the qubits used in 
Ref.~\cite{Riste2013} have charge noise fluctuations of the
same order as the telegraph noise measured here, but charge noise of that 
magnitude is expected, as explained by
the different parameter range of those qubits: $\omega_{01}/2\pi = 4.387$ GHz
and $\eta/2\pi = -334$ MHz, giving $\epsilon_1 \approx 2$ MHz.

\section{Calculation of $\Omega_{ZZ}$}
\label{app:omegaZZ}

Two capacitively coupled qubits have an XX-type coupling of the form $g(\ket{01}\bra{10} +
\ket{10}\bra{01})$, where the coupling constant $g$ is half the swap rate between the qubits.
The interaction between the higher levels, $\sqrt{2}g(\ket{11}\bra{20} + \ket{02}\bra{11}) +
\sqrt{2}g(\ket{11}\bra{20} + \ket{02}\bra{11})$, results in a repulsion of the $\ket{11}$
level from the $\ket{02}$ and $\ket{20}$ levels; this energy shift in the $\ket{11}$ level
produces a ZZ-type interaction between the qubits. In the far-detuned limit, neglecting the
XX-coupling, the two-qubit Hamiltonian becomes
\begin{align}
  & H =  \omega_1\ket{10}\bra{10} +  \omega_2\ket{01}\bra{01} \nonumber \\
& \hspace{0.6cm}      + \left(\omega_1 + \omega_2 + \Omega_{ZZ}\right)\ket{11}\bra{11},        \\
  & \Omega_{ZZ} = \frac{2g^2}{\Delta - \eta_2} + \frac{2g^2}{-\Delta - \eta_1},
\end{align}
where $\omega_n$ and $\eta_n$ are the qubit frequencies and nonlinearities, respectively, and
$\Delta = \omega_1 - \omega_2$. In our system, $\eta_1 = \eta_2 \equiv \eta$, giving
\begin{equation} \label{eq:omegazz}
 \Omega_{ZZ} = \frac{4g^2 \eta}{\Delta^2 - \eta^2}.
\end{equation}
When both qubits are simultaneously performing an RB sequence, phase error
$\phi$ per idle gate in qubit A is
\begin{equation} \label{eq:rb2qubitPhaseError}
 \phi = \pm \frac{\Omega_{ZZ}}{2} t_\text{gate}
\end{equation}
where $t_\text{gate}$ is the idle gate duration, and the frequency shift $\pm \Omega_{ZZ}/2$ assumes centering the qubit frequency.
This gives $\braket{\phi^2} = (\Omega_{ZZ}t_\text{gate})^2/4$, and since for RB the error per gate is $E = \braket{\phi^2}/6$ [see Eq.\ (\ref{eq:rbErrorDerivation})], we arrive at Eq.~(\ref{eq:omegaZZtheory}) for the error per gate due to the $\Omega_{ZZ}$ interaction,
\begin{equation} \label{eq:omegazzError}
 E = \frac{\pi^2}{6}\left( \frac{\Omega_{ZZ}}{2\pi} t_\text{gate}\right)^2.
\end{equation}

\newpage

\section{Fits to Gate Errors in Figure \ref{fig:gates}}
\label{app:gatesFits}

For the data in the Fig.~\ref{fig:gates}, the fits are made 
either to a simple linear model in the case of Markovian noise (the $XX$ and 
$YX$ cases) or to a quadratic and linear model in the case of non-Markovian 
noise (the $I$ and $Z$ cases). There is no offset in
any fit.
\begin{center}
 \begin{tabular}{c|c|c}

  Gate & Linear Term & Quadratic Term \\
  & ($10^{-6}$ error/ns) & ($10^{-6}$ error/ns$^2$) \\ \hline
  $I$ & 17 & 0.22     \\ \hline
  $XX$ & 20 & -       \\ \hline
  $Z$ & 24 & 0.18     \\ \hline
  $YX$ & 22 & -

 \end{tabular}
\end{center}
Note that the contribution from $T_1 = 26.7 \, \mu$s to the linear portion of the error, given
by Eq.~\ref{eq:rbT1}, is $9.3\times10^{-6}$ error/ns, or roughly half of the error measured.
The remainder is equivalent to a white noise dephasing with time constant $T_{\text{white}}
\approx 30 \, \mu$s, according to Eqs.~\ref{eq:rbErrorDerivation} and \ref{eq:whiteNoise}. The
quadratic terms correspond with $T_{\phi2} \approx 1 \, \mu$s.

\newpage

\section{Telegraph Noise Measured in Other Devices}
\label{app:otherDevices}

\begin{figure}[th]
  \centering
  \includegraphics[width=0.48\textwidth]{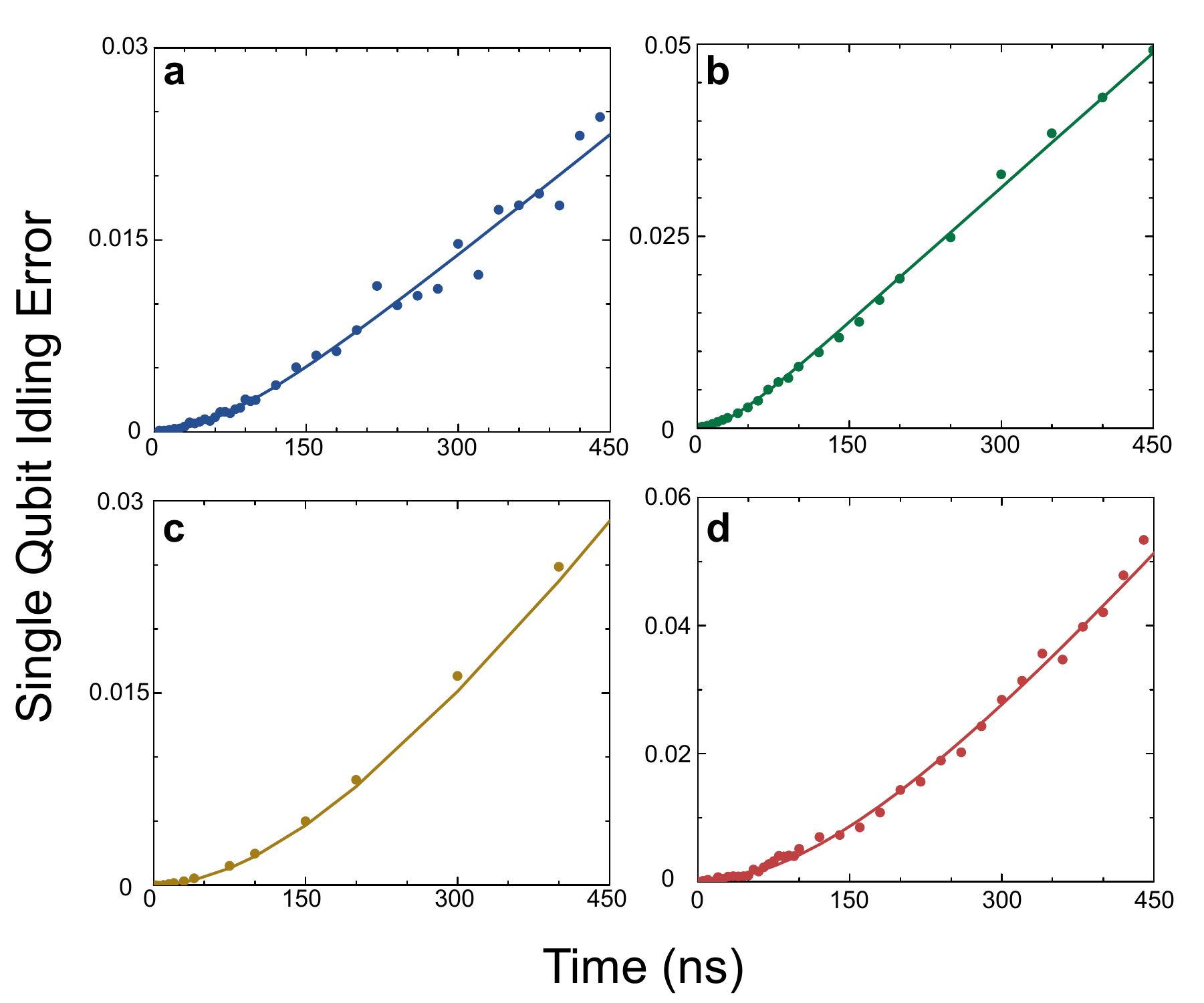}
  \caption{(color online) Telegraph noise measured with RB Ramsey in other devices.
  All fits included $T_1$ and telegraph noise only (Eq.~(\ref{eq:telegraphFit})).
  (a) A reproduction of Fig.~\ref{fig:rbRamsey} data for reference.
  (b) Measurement of another Xmon on the same chip as the device.
  (c) Measurement of an Xmon qubit from another sample; see \cite{Barends2014} for device details.
  (d) Measurement of a gmon qubit; see \cite{Neill2015} for device details.
  }
  \label{fig:otherDevices}
\end{figure}

\begin{table}[bh]
 \begin{center}
 \begin{tabular}{c|c|c|c|c|c|c}

   Sample & $f_{10}$ & $df/d\Phi$ & $T_1$ & $T_{\text{sw}}$ & $\Delta f_{10}$ & Device \\
  (see text) & (GHz) & (GHz/$\Phi_0$) & ($\mu$s) & (ns) & (kHz) & Details \\ \hline
  a & 4.9 & 4.81 & 26.7 &  84 & 479 & q2 of \cite{Kelly2015} \\ \hline
  b & 4.8 & 5.36 & 15.7 & 183 & 274 & q0 of \cite{Kelly2015} \\ \hline
  c & 5.5 & 3.96 & 22.2 & 201 & 199 & q2 of \cite{Barends2014} \\ \hline
  d & 4.9 & 6.62 & 15.7 &  32 & 528 & q1 of \cite{Neill2015}

 \end{tabular}
 \end{center}
 
 \caption{Fits for telegraph noise measured in other devices (Fig.~\ref{fig:otherDevices}); see text and references for sample details.}
 \label{tab:otherDevices}
\end{table}

Telegraph noise has been observed in many other devices.
In Fig.~\ref{fig:otherDevices}, we present RB Ramsey measurements of three other devices that show telegraph noise, with the data from Fig.~\ref{fig:rbRamsey} reproduced for reference (a);
one is another device on the same chip (b), one another Xmon with different parameters \cite{Barends2014} (c), and the last a gmon qubit \cite{Neill2015} (d).
All fits were to $T_1$ and telegraph noise only, Eq.~(\ref{eq:telegraphFit}), with fit parameters given in Table~\ref{tab:otherDevices}.

\addcontentsline{toc}{chapter}{Bibliography}
\bibliography{main}

\end{document}